\documentclass[aps,pra,twocolumn,superscriptaddress,amsmath,amssymb,a4paper,showpacs]{revtex4} 
\usepackage{graphicx}
\usepackage{amsmath}
\usepackage{amsfonts}
\usepackage{amssymb}
\usepackage{hyperref}

\begin{document}

\title{Phase transitions and elementary excitations in spin-1 Bose gases with Raman-induced spin-orbit coupling}

\author{Zeng-Qiang Yu}
\email{zqyu.physics@outlook.com}
\affiliation{Institute of Theoretical Physics, Shanxi University, Taiyuan 030006, China}

\begin{abstract}
  We study the ground state phase diagram and the quantum phase transitions in spin-1 Bose gases with Raman induced spin-orbit coupling. In addition to the Bose-Einstein condensates with uniform density, three types of stripe condensation phases that simultaneously break the U(1) symmetry and the translation symmetry are identified. The transitions between these phases are investigated, and the occurrences of the various tricritical points are predicted. The excitation spectra in the plane-wave phase and the zero-momentum phase show rich roton-maxon structures, and their instabilities indicate the tendency to develop the crystalline order. We propose the atomic gas of $^{23}$Na could be a candidate for observing the stripe condensate with high contrast fringes.
\end{abstract}

\maketitle

\section{Introduction}

The remarkable realization of synthetic spin-orbit (SO) coupling in quantum gases generates great interest recently and is opening up new perspective in exploring the many-body phenomena with ultracold atoms~\cite{Review}. So far, a specific type of SO coupling, which is induced by a pair of Raman laser beams, has been experimentally achieved in atomic gases of $^{87}$Rb~\cite{NIST2011}, $^{40}$K~\cite{SXU2012}, and $^{6}$Li~\cite{MIT2012}. In contrast to solid-state materials, ultracold atoms provide a unique platform to study the rich SO effects in bosons. Previously, the interesting properties of the SO coupled spin-half Bose-Einstein condensates have been extensively studied on both the experimental side~\cite{NIST2011,USTC2012,NIST2013,USTC2014,WSU2014-1,WSU2014-2,USTC2015} and the theoretical side~\cite{Ho,Trento1,Trento2,Trento3,Zhengwei,Yu2014}. Very recently, the Raman coupling scheme is experimentally generalized to spin-1 bosons by NIST group~\cite{NIST2015}, and novel condensation phases in this system are also theoretically predicted~\cite{Lan2014,Natu2015}.

The purpose of the present work is to study the interactions effects on the quantum phase transitions in the spin-1 SO coupled Bose gases. Our main results are summarized in Fig.~1, which shows the ground-state phase diagrams in terms of the Raman coupling strength $\Omega$ and the quadratic Zeeman field $\Lambda$. Three types of stripe condensation phases that simultaneously break the U(1) symmetry and the translation symmetry are identified with either antiferromagnetic interaction or ferromagnetic interaction. These stripe phases, as well as the plane-wave (PW) phase and the zero-momentum (ZM) phase, are characterized by the spin magnetization and the crystalline order. The occurrences of the various tricritical points are also predicted.

\begin{figure}[b!] 
\includegraphics[width=7.4cm]{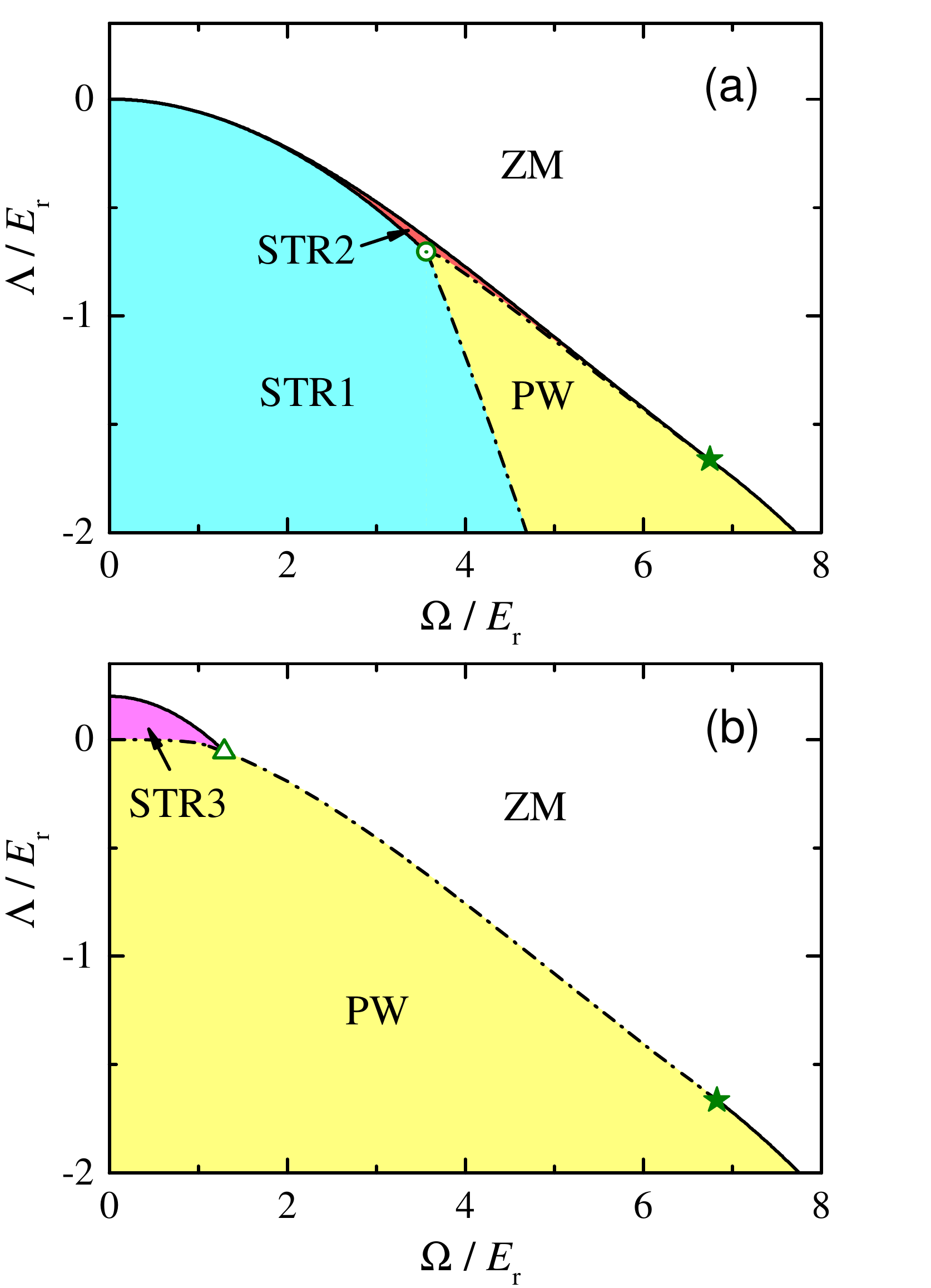}
\caption{Ground-state phase diagram of SO coupled spin-1 Bose gases with (a) antiferromagnetic interaction and (b) ferromagnetic interaction. Three types of stripe condensation phases are labeled as STR1, STR2 and STR3, respectively. The quantum phase transitions could be either first order (dash-dotted line) or second order (solid line). Various tricritical points are indicated by the symbols $\bigstar$, $\odot$, and $\bigtriangleup$. Interactions parameters: (a) $c_2\bar n=0.1E_{\rm r}$ and (b) $c_2\bar n=-0.1E_{\rm r}$; for both plots $c_0\bar n=1E_{\rm r}$.}
\end{figure}

The paper is organized as follows. In the next section, we first recall the unique feature of the single-particle dispersion in this spin-1 system and interpret the magnetic tricritical transition in the noninteracting limit. Then, the effects of many-body interactions on the magnetic phase transition are considered in Sec.~III, and the elementary excitations in the PW phase and the ZM phase are also studied. In Sec.~IV, the ground-state phase diagram are determined via a variational approach, and the stripe phases are characterized by the spatial modulations of density and magnetization. Finally, in Sec.~V, we conclude with discussions on the experimental relevance of our theory. For completeness, the calculation details are given in Appendixes.

\section{Single-Particle Physics}

We consider the experimental scheme that three Raman laser beams propagating along $x$ direction are employed to generate SO coupling in spin-1 bosons~\cite{NIST2015}. In this setup, the momentum transfer between the adjacent hyperfine states is achieved via the two-photon Raman transition. The effective Hamiltonian (in the laboratory frame) for single atom is given by (set $\hbar=1$)
\begin{align}
  \tilde H_{\rm R} = \frac{\vec p^2}{2m} + \frac{\Omega}{\sqrt{2}} \left( F_x\cos 2k_{\rm r} x  + F_y \sin 2k_{\rm r}x \right) + \Lambda F_z^2,
  \label{H0_lab}
\end{align}
where $k_{\rm r}$ is the recoil momentum of the Raman lasers, $\{F_x, F_y, F_z\}$ are Pauli matrices for spin-1, $\Omega$ is the Raman coupling strength, and $\Lambda$ is the quadratic Zeeman field~\cite{note1}. By applying the unitary transformation, $H_{\rm R}=U^\dag \tilde H_{\rm R} U$ with $U=e^{-2ik_{\rm r}x F_z}$,
one can remove the coordinate dependence of Hamiltonian (\ref{H0_lab}).
In the rotating frame, the single-particle Hamiltonian is rewritten as
\begin{align}
  H_{\rm R} = \epsilon_p - \tfrac{2}{m} k_{\rm r}p_x F_z  + \tfrac{1}{\sqrt{2}} \Omega F_x + \Lambda' F_z^2,
  \label{H0}
\end{align}
with $\epsilon_p= p^2/(2m)$, $\Lambda'=\Lambda+4E_{\rm r}$, and $E_{\rm r}=k_{\rm r}^2/(2m)$. The second term of the Hamiltonian (\ref{H0}) represents a linear coupling between spin and momentum.

We choose momentum as a good quantum number to determine the single-particle ground state~\cite{note_singleparticle}. The dispersion relation of lowest band is given by
\begin{align}
  \varepsilon_{\bf p} =  \epsilon_p + \tfrac{2}{3}\Lambda'  - 2\sqrt{A_{p_x}} \cos \tau_{p_x},
\end{align}
with $\tau_{p_x}=\tfrac{1}{3} \arccos (B_{p_x}/A_{p_x}^{3/2})$,  $ A_{p_x} = \tfrac{1}{9}\Lambda'^2 + \tfrac{1}{6} \Omega^2 + \tfrac{16}{3} E_{\rm r} \epsilon_{p_x}$, and $B_{p_x}= \tfrac{1}{27}\Lambda'^3 + \tfrac{1}{12} \Lambda'(\Omega^2-64 E_{\rm r} \epsilon_{p_x}) $. The energy eigen-state for the same band is given by
\begin{align}
  \phi_{\vec p}(\vec r)
  = \sqrt{1 \over V} \begin{pmatrix}
    a_+ \\ a_0 \\ a_-
  \end{pmatrix} e^{i \vec p\cdot \vec r},
\end{align}
with $a_\pm=-a_0\Omega/C_\pm$, $a_0^2=[1+\Omega^2/C_+^2+ \Omega^2/C_-^2]^{-1}$, and $C_\pm = \tfrac{2}{3}\Lambda' \mp \tfrac{4}{m}p_xk_{\rm r} + 4\sqrt{A_{p_x}}\cos\tau_{p_x}$. For $p_x=0$, $a_+=a_-$.

As first pointed out by Lan and \"{O}hberg~\cite{Lan2014}, the single-particle ground state can be dramatically tuned by the Raman coupling strength and the quadratic Zeeman field (see Fig.~2). When $\Lambda$ is large enough, $\varepsilon_{p_x}$ has only one minimum, and the ground state is the ZM state; when $\Lambda\rightarrow-\infty$, the lowest energy occurs at two opposite momenta, and the ground state is the PW state with two-fold degeneracy. In the intermediate region, there are three local energy minima for $\Omega$ below a critical value, and the ground state is determined by the competition between them. The three minima merge together at the tricritical point $(\Omega_\star,\Lambda_\star)$, where the spectrum becomes a extremely flat band in $x$ direction
\begin{align}
  \varepsilon_{p_x} = \varepsilon_\star + \lambda_{\star} p_x^6  \qquad (p_x\rightarrow 0),
\end{align}
with $\varepsilon_\star = 16(2 - \sqrt{5})E_{\rm r}$ and $\lambda_{\star}=\big[(256\sqrt{5}-511)mk_{\rm r}^4\big]^{-1}$. For $\Omega<\Omega_\star$, the transition between the PW state and the ZM state is first order, and the ground-state momentum exhibits a sudden jump at the transition. For $\Omega>\Omega_\star$, the PW-ZM transition is second order, and the ground-state wave function evolves continuously across the phase boundary.

\begin{figure}[t!] 
\includegraphics[width=7.9cm]{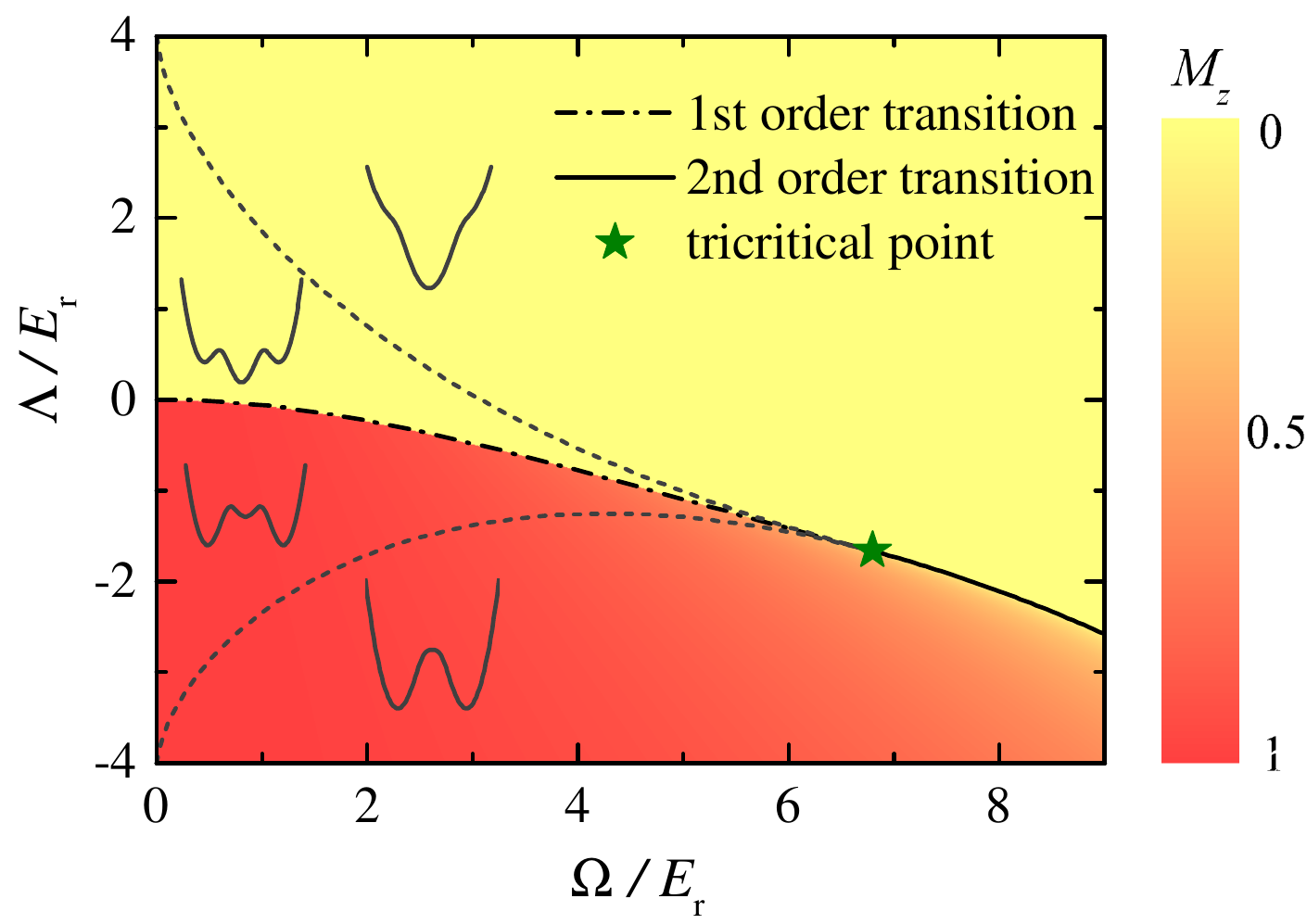}
\caption{Ground-state phase diagram of a spin-1 boson with Raman coupling. The insets show schematic dispersion relation of the lowest band in different regions, where the spectrum has one, two, and three minima, respectively. }
\end{figure}

To further get a quantitative description of the tricritical transition, we expand the single-particle spectrum in the vicinity of $p_x=0$. Up to the sixth order, we have
\begin{align}
  \varepsilon_{p_x} = \varepsilon_0 + \lambda_2 p_x^2 + \lambda_4 p_x^4 + \lambda_6 p_x^6 + \mathcal{O} (p_x^8),
  \label{Ep_expansion}
\end{align}
where $\varepsilon_0=\tfrac{1}{2}(\Lambda'-\sqrt{\Lambda'^2+2\Omega^2})$, and explicit form of coefficients $\lambda_2, \lambda_4$ and $\lambda_6$ are given in Appendix~A. The dispersion relation in Eq.~(\ref{Ep_expansion}) is reminiscent to the Landau's phase transition theory~\cite{textbook}, in which the free energy is expanded in terms of the order parameter. With an analogous analysis, we find the conditions for the PW-ZM transition.
When $\lambda_4^2=4\lambda_2 \lambda_6$ (with $\lambda_4<0$ and $\lambda_2,\lambda_6>0$), a first-order transition takes place; when $\lambda_2=0$ (with $\lambda_4, \lambda_6>0$), a second-order transition happens. The first-order transition and the second-order transition meet at the tricritical point $(\Omega_\star,\Lambda_\star)$, where the condition
\begin{align}
  \lambda_2=\lambda_4=0 \label{tricritical_condition}
\end{align}
is satisfied. Although the dispersion in Eq.~(\ref{Ep_expansion}) is approximate, it provides an accurate description of the low-energy physics in the vicinity of the tricritical point. From the condition (\ref{tricritical_condition}), we find the tricritical point at
\begin{align}
  \Omega_\star &= 16\sqrt{5\sqrt{5}-11} \, E_{\rm r} \simeq 6.7946 E_{\rm r}, \\
  \Lambda_\star &= 4 \big(13-6\sqrt{5} \big) \, E_{\rm r} \simeq -1.6656 E_{\rm r}.
\end{align}

Due to the spin-momentum coupling, when the atom's momentum varies, its spin polarization also changes. The momentum of the single-particle ground state $p_0$ satisfies the stationary condition
\begin{align}
  \partial_{p_x} \varepsilon_{p} \big|_{p=p_0} = 0.
\end{align}
Applying the Hellmann-Feynman theorem to the left-hand side above, we find
\begin{align}
  p_{0} = 2k_{\rm r} M_z,    \label{P-M_relation}
\end{align}
where $M_z=\langle F_z \rangle$ is the longitudinal magnetization. Relation (\ref{P-M_relation}) implies an accompanied magnetic transition across the PW-ZM phase boundary.
$M_z$ is finite in the PW state and is zero in the ZM state. When a first-order transition between these two phases happens, a sudden jump in the magnetization can be observed; when a second-order transition takes place, the nonzero magnetization emerges continuously in the PW state.
The landscape of the magnetization is shown in the phase diagram of Fig.~2. In the PW phase regime, the ground state has two fold degeneracy, $M_z$ is chosen to be positive with a spontaneous breaking of the $Z_2$ symmetry.

\section{Condensate with uniform density}

\subsection{PW condensate and ZM condensate}

Now we turn to the interactions effects in a many-body system.

The interactions between spin-1 bosons include two parts: the density-density interaction and the spin-spin interaction. The interaction Hamiltonian reads
\begin{align}
  H_{\rm int} = \frac{c_0}{2} \int d \vec r \,  n^2(\vec r) + \frac{c_2}{2} \int d \vec r \,  \mathcal{\vec F} (\vec r)\cdot \mathcal{\vec F}(\vec r),
\end{align}
where $n(\vec r)= \psi^\dag \psi$ is the local density operator, $\mathcal{\vec F} (\vec r)= (\mathcal{F}_x, \mathcal{F}_y, \mathcal{F}_z)$ is the local spin operator with $\mathcal{F}_{x,y,z} \equiv \psi^\dag F_{x,y,z} \psi$, $\psi^\dag= (\psi^\dag_{+},\psi_0^\dag,\psi_{-}^\dag)$ is the creation operator for the spin-1 atoms, and $c_0$ and $c_2$ are interaction strength parameters. In this work, the density-density interaction is assumed to be repulsive $(c_0>0)$, and the spin-spin interaction could be either antiferromagnetic $(c_2>0)$ or ferromagnetic $(c_2<0)$.

In analogy to the single-particle case, we first assume the many-body ground state is a Bose-Einstein condensate with single momentum macroscopically occupied. The condensate wave function $\varphi\equiv \langle \psi \rangle$ is written as
\begin{align}
    \varphi(\vec r) = \sqrt{\bar n} \begin{pmatrix}
      \alpha_+ \\ \alpha_0 \\ \alpha_-
    \end{pmatrix} e^{i \vec p_c \cdot \vec r}, \label{condensate_PW}
\end{align}
where $\bar n$ is the average density of atoms, $\vec p_{\rm c} = (p_{\rm c},0,0)$ is the condensation momentum, and $\alpha_{\pm,0}$ are real parameters satisfying the normalization condition $\alpha_+^2+\alpha_0^2+\alpha_-^2=1$. In general, the condensate wave function does not necessarily coincide with the single-particle ground state.

At mean-field level, we replace the field operator $\psi$ by the condensate wavefunction $\varphi$ and write the energy functional of the system as
\begin{align}
  \mathcal{E} = \int d\vec r\, \Big[ \varphi^\dag H_{\rm R} \varphi + \frac{c_0}{2} \left( \varphi^\dag \varphi \right)^2 + \frac{c_2}{2} \sum_{j=x,y,z} \left( \varphi^\dag F_j \varphi \right)^2 \Big].
  \label{energy_functional}
\end{align}
The condensate wavefunction should minimize the total energy. From the stationary condition $\partial_{p_{\rm c}} \mathcal{E}=0$ and the wavefunction ansatz in (\ref{condensate_PW}), we find a simple relation between the condensation momentum and the longitudinal magnetization,
\begin{align}
  p_{c} = 2k_{\rm r} M_z,   \label{P-M_relation2}
\end{align}
with $M_z=\alpha_+^2-\alpha_-^2$. In contrast with the single-particle result in Eq.~(\ref{P-M_relation}), the interactions effects have been included in this relation, where $p_{\rm c}$ in principle could be different from $p_0$. The spinor parameters $\alpha_{\pm,0}$ satisfy the time-independent Gross-Pitaevskii (GP) equation,
\begin{align}
  \mathcal{L} \begin{pmatrix}
    \alpha_+ \\ \alpha_0 \\ \alpha_-
  \end{pmatrix}
  = \mu \begin{pmatrix}
    \alpha_+ \\ \alpha_0 \\ \alpha_-
  \end{pmatrix},
  \label{GP_equation}
\end{align}
where $\mu$ is the chemical potential, and
$$\mathcal{L} = H_{\rm R}(p_{\rm c}) + c_0\bar n + c_2\bar n \left[ M_zF_z + \sqrt{2}\alpha_0(\alpha_+ + \alpha_-) F_x \right].$$
Once $p_{\rm c}$ and $\alpha_{\pm,0}$ are determined from Eqs.~(\ref{P-M_relation2}) and (\ref{GP_equation}), the equation of state for the PW phase and ZM phase thus can be obtained.

\begin{figure}[t!] 
\includegraphics[width=7.5cm]{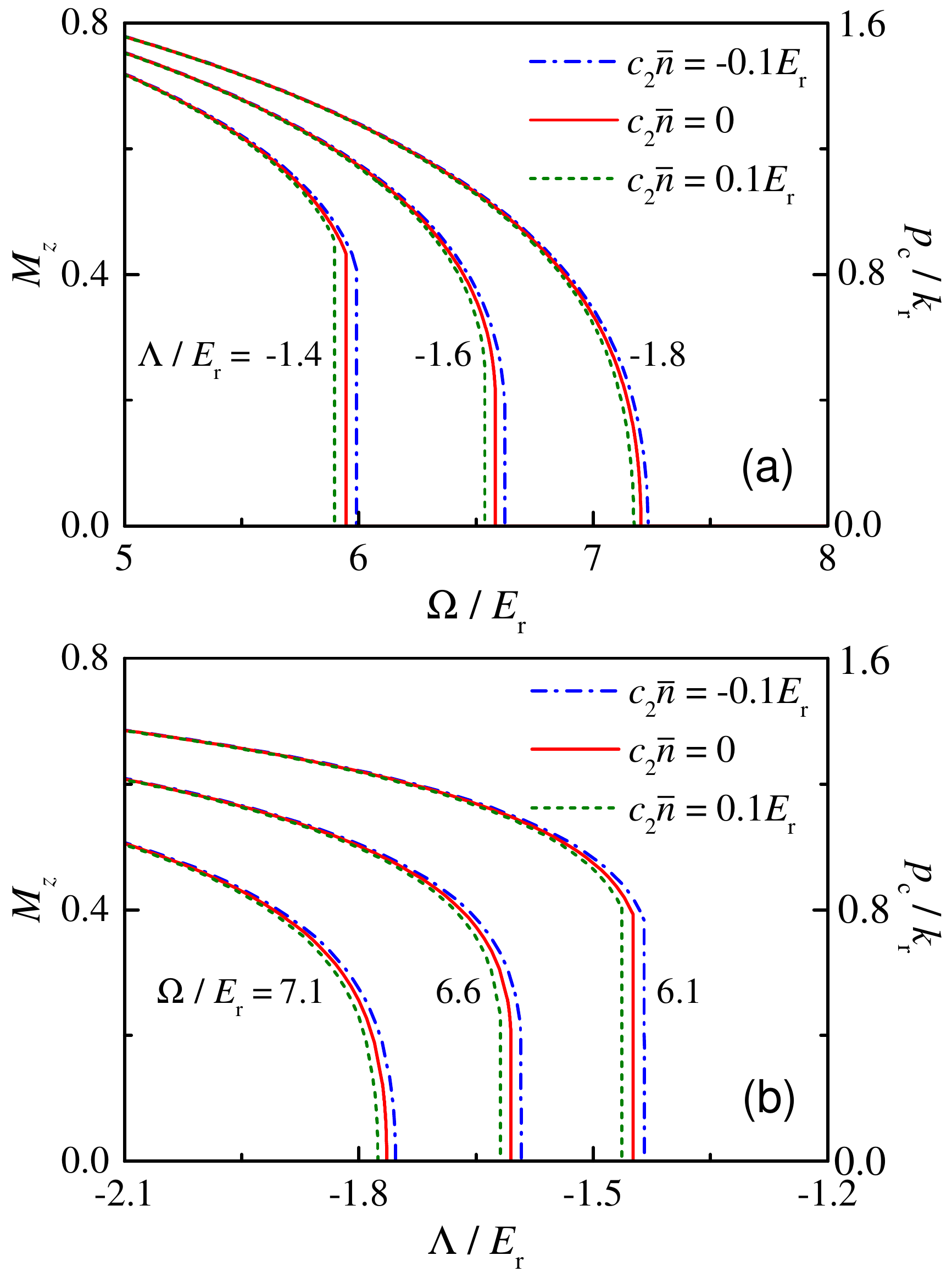}
\caption{Magnetic phase transitions between the PW phase and the ZM phase for different spin-dependent interaction strength $c_2$. The condensation momentum is related to the spin magnetization via the relation $p_{\rm c}=2k_{\rm r}M_z$ (see the right axes). For both plots, $c_0\bar n=1E_{\rm r}$.}
\end{figure}

When the interactions between the atoms are spin-independent, i.e., $c_2=0$, the solution of GP equation (\ref{GP_equation}) reduces to the single-particle wave function with $\mu=\varepsilon_{p_{\rm c}}+c_0\bar n$, and the ground-state phase diagram remains the same as the noninteracting case (Fig.~2). When the interactions become spin-dependent, the condensate wave function is different from the single-particle state and, consequently, the transition between the PW phase and the ZM phase is affected. As shown in Fig.~3, for $c_2<0$ ($c_2>0$), the phase boundary moves toward the side with larger (smaller) $\Omega$ and $\Lambda$, and the discontinuity magnitude of $M_z$ decreases (increases) at the first-order transition~\cite{note_transition}. When the spin-spin interaction is weak, such corrections are typically small.

\subsection{Elementary excitations}

\begin{figure} 
\includegraphics[width=7.4cm]{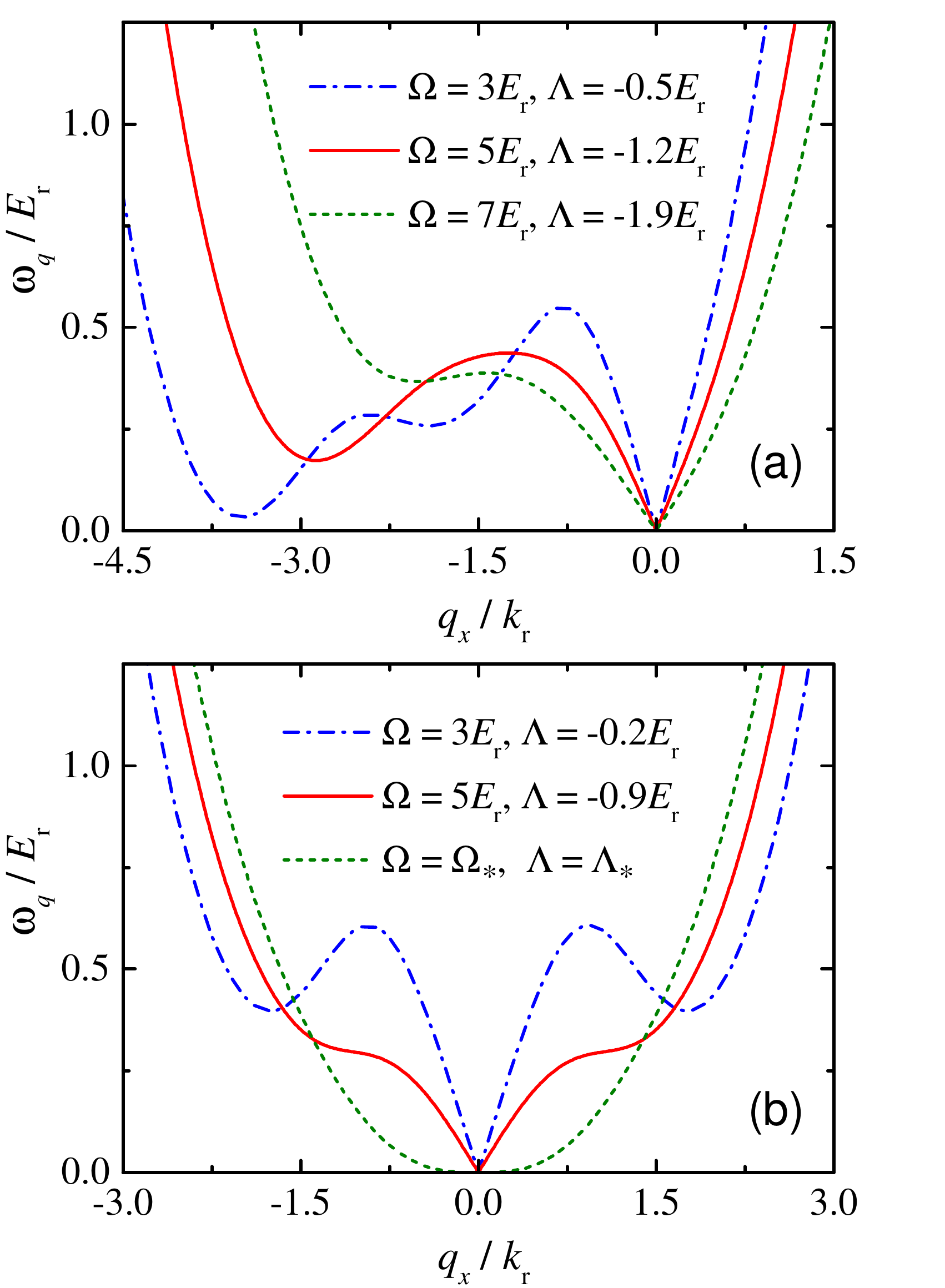}
\caption{Lowest band of elementary excitations in (a) the PW phase and (b) the ZM phase for different values of $\Omega$ and $\Lambda$. For both plots, $c_0\bar n=1E_r$ and $c_2=0$.}
\end{figure}

Beyond the mean-field level, one can use Bogoliubov theory to take account of quantum fluctuations. In the PW phase and the ZM phase, the grand canonical Hamiltonian for the fluctuation part is given by
\begin{align}
  \mathcal{K}_{\rm Bog} =\sum_{q_x>0} \left\{ \Psi_{\vec q}^\dag \mathbb{K}_{\vec q} \Psi_{\vec q} -  {\rm Tr} \left[ K_{\rm R}(\vec p_{\rm c}- \vec q) + \Sigma_{\rm N} \right] \right\}
  \label{H_Bogoliubov}
\end{align}
with $K_{\rm R}(\vec p)=H_{\rm R}(\vec p)-\mu$, and
\begin{align}
  \mathbb{K}_{\vec q} = \begin{bmatrix}
    K_{\rm R}(\vec p_{\rm c} + \vec q)  + \Sigma_{\rm N} & \Sigma_{\rm A} \\
    \Sigma_{\rm A} & K_{\rm R}(\vec p_{\rm c} - \vec q)  + \Sigma_{\rm N}
  \end{bmatrix}.
\end{align}
Here, $\Psi_{\vec q}^\dag$ is the creation operator in the Beliaev representation with $\vec q$ being the excitation momentum, $\Sigma_{\rm N}$ and $\Sigma_{\rm A}$ are the normal self-energy and anomalous self-energy, respectively, whose explicit expressions are given in Appendix B. The quadratic Hamiltonian (\ref{H_Bogoliubov}) can be solved via the Bogoliubov transformation, and the elementary excitations are readily determined from
\begin{align}
  {\rm Det} \begin{bmatrix}
    K_{\rm R}(\vec p_{\rm c} +  \vec q)  + \Sigma_{\rm N}-\omega & \Sigma_{\rm A} \\
    -\Sigma_{\rm A} & \!-K_{\rm R}(\vec p_{\rm } - \vec q)  - \Sigma_{\rm N} - \omega
  \end{bmatrix} = 0.
\end{align}

Due to the emergence of the off-diagonal long-range order, the lowest branch excitations in the long wavelength limit is the gapless phonon mode. The gapless feature of the spectrum is mathematically guaranteed by a modified version of the Hugenholtz-Pines relation~\cite{HP-relation},
\begin{align}
   {\rm Det}\left[ H_{\rm R}(p_{\rm c}) - \mu + \Sigma_{\rm N}(0,0) - \Sigma_{\rm A}(0,0) \right] = 0,  \label{HP_relation}
\end{align}
which is indeed satisfied at the Bogoliubov level. Across the first-order transition between the PW phase and the ZM phase, the sound velocity of the phonon mode exhibits a sudden jump at the phase boundary. At the second-order PW-ZM transition, sound velocity vanishes (in $x$ direction), which is similar to the spin-half case~\cite{Trento2,Zhengwei}. At the tricritical point $(\Omega_\star,\Lambda_\star)$, the phonon mode is extremely soft and shows a novel cubic dispersion in long-wavelength limit,
\begin{align}
  \omega_{q_x} = \eta |q_x|^3, \qquad (q_x\rightarrow 0).
\end{align}
For $c_2=0$, we find $\eta=\sqrt{c_0n E_{\rm r}/(2\sqrt{5}-4)}/(8k_{\rm r}^3)$.

At larger $q_x$, the lowest band of excitations could show rich roton-maxon structures, as plotted in Fig.~4. Such a nonmonotonic dispersion relation attributes to the triple-well/double-well structure of the single-particle spectrum. In the ZM phase, the roton-maxon structure (if it exists) is symmetric, and the spectrum has two degenerate local minima at opposite momenta; in the PW phase, the roton-maxon structure is asymmetric, and the spectrum has either one or two local minima at $q_x<0$. Previously, the roton excitations in the two-component SO coupled Bose gases have been experimentally observed via the Bragg spectroscopy measurement~\cite{USTC2015,WSU2014-2}; a similar technique can be employed in the spin-1 system to detect the double-roton structure (see Appendix~C for details).

While the excitations spectra are always stable for $c_2=0$, the spin-dependent interaction (being either antiferromagnetic or ferromagnetic) could cause various instabilities, as shown in Fig.~5.
The onset of the occurrence of the instability is corresponding to the situation that the roton gap vanishes. In the unstable region, the ZM phase suffers the dynamic instability that the energies of  the roton excitations become imaginary, and the PW phase suffers the energetic instability that the energies of roton excitations become negative. When the dispersion shows an asymmetric double-roton structure, the energetic instability occurs near the roton with a lager (smaller) $|q_x|$ for $c_2>0$ ($c_2<0$). Similar to the spin-half case~\cite{USTC2015}, the closure of the roton gap indicates the tendency to develop crystalline order in the system. The different type of instabilities imply the existence of stripe phases with different crystalline orders.

\begin{figure}[t!] 
\includegraphics[width=7.4cm]{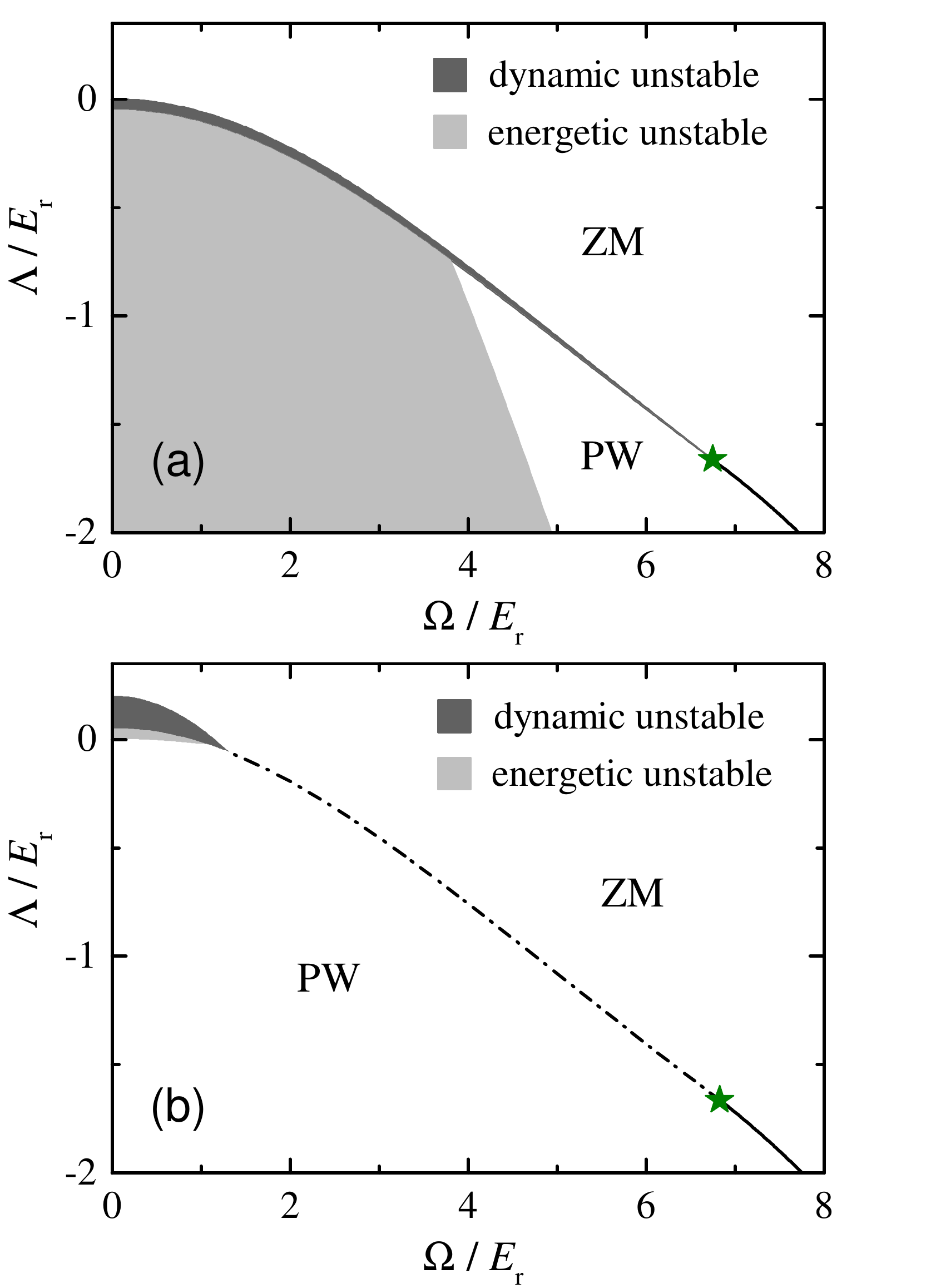}
\caption{Stability diagram of the PW phase and the ZM phase with (a) antiferromagnetic interaction and (b) ferromagnetic interaction. The roton excitations suffer the dynamic (energetic) instability in the region with dark-gray (light-gray) shadow. Interactions parameters are the same as Fig.~1.}
\end{figure}

\section{Stipe Condensates}

\subsection{Three types of stripe condensates}

To include the stripe phases, we employ the variational method and write the condensate wavefunction as
\begin{align}
  \varphi=\sqrt{\bar n} \left[ \xi_+ \!
  \begin{pmatrix}
    \alpha_+ \\ \alpha_0 \\ \alpha_-
  \end{pmatrix} \! e^{ip_{\rm c}x}
   + \xi_0 \!
   \begin{pmatrix}
    \beta_1 \\ \beta_0 \\ \beta_1
  \end{pmatrix}
   + \xi_- \! \begin{pmatrix}
    \alpha_- \\ \alpha_0 \\ \alpha_+
  \end{pmatrix} \! e^{-ip_{\rm c}x} \right]
  \label{variation_wf},
\end{align}
with $p_{\rm c}\!\neq \!0$.
Here, $\alpha_{\pm,0}$ and $\beta_{0,1}$ are real parameters describing the spinor configurations with different momenta, and $\xi_{\pm,0}$ are complex coefficients for each components. These parameters satisfy the normalization constraints $\alpha_+^2+\alpha_0^2+\alpha_-^2=1$, $\beta_0^2+2\beta_1^2=1$ and $|\xi_+|^2+|\xi_0|^2+|\xi_-|^2=1$.
When $\xi_0=\xi_-=0$, the variation ansatz (\ref{variation_wf}) recovers the wavefunction of the PW phase; when $\xi_\pm=0$, it describes the ZM condensate; when two or all of the coefficients $\{\xi_0,\xi_{\pm}\}$ are nonzero, it represents a stripe phase that breaks the translation symmetry.

According to Eq.~(\ref{energy_functional}), the variation energy for the ground state can be written in terms of the parameters $\alpha_{\pm,0}$, $\beta_{0,1}$, $p_{\rm c}$, $|\xi_{\pm ,0}|$, and $\theta\equiv \arg\xi_0-(\arg\xi_+ + \arg \xi_-)/2$. Through a straightforward numeric minimization with respect to the variation parameters, we find the ground-state phase diagram as shown in Fig.~1. Besides the PW phase and the ZM phase, three types of stripe phases are identified as follows.

(i) The STR1 phase with $\xi_0=0$ and $|\xi_+|=|\xi_-|=1/\sqrt{2}$. In this phase, the longitudinal magnetization $\mathcal{F}_z$ vanishes everywhere, and density distribution shows a spatial modulation with a period of $\pi/p_{\rm c}$,
\begin{align}
  n(\vec r) = \bar n \left[ 1+ 2 \left|\xi_+\xi_-\right| \left(\alpha_0^2+2\alpha_+\alpha_-\right) \cos(2p_{\rm c}x+2\theta')\right], \label{density_modulation1}
\end{align}
where $\theta'\equiv (\arg \xi_+ - \arg\xi_-)/2$ determines the positions of the peaks in the density profile.

(ii) The STR2 phase with $\xi_0\neq 0$, $|\xi_+|=|\xi_-|\neq 0$, and $\theta=\pi/2$. In this phase, both density and spin polarization are nonuniform. The density distribution is in the same form as Eq.~(\ref{density_modulation1}), and longitudinal magnetization oscillates with a period being twice of the density modulation
\begin{align}
  \mathcal{F}_z(\vec r)  = \bar n \sqrt{2} \left|\xi_0\xi_+\right| (\alpha_+ - \alpha_-) \beta_1 \sin(p_{\rm c}x + \theta').
\end{align}
The relative phase between the density and spin modulation is fixed, and a node of $\mathcal{F}_z(\vec r)$ is always corresponding to a peak of $n(\vec r)$.

(iii) The STR3 phase with $\xi_0\neq 0$, $|\xi_+|=|\xi_-|\neq 0$, and $\theta=0$. In this phase, $\mathcal{F}_z=0$, and the condensate density oscillates with a period of $2\pi/p_{\rm c}$,
\begin{align}
  n(\vec r) &= \bar n \Big[ 1 + 2\left|\xi_+\xi_-\right| \left(\alpha_0^2+2\alpha_+\alpha_-\right) \cos(2p_{\rm c}x+2\theta')
  \nonumber \\
  & \quad \left. + 4\left|\xi_0\xi_+\right| \left( \alpha_0\beta_0 + \alpha_+ \beta_1 + \alpha_-\beta_1 \right) \cos(p_{\rm c}x+\theta') \right].
\end{align}

\begin{figure}[t!] 
\includegraphics[width=7.9cm]{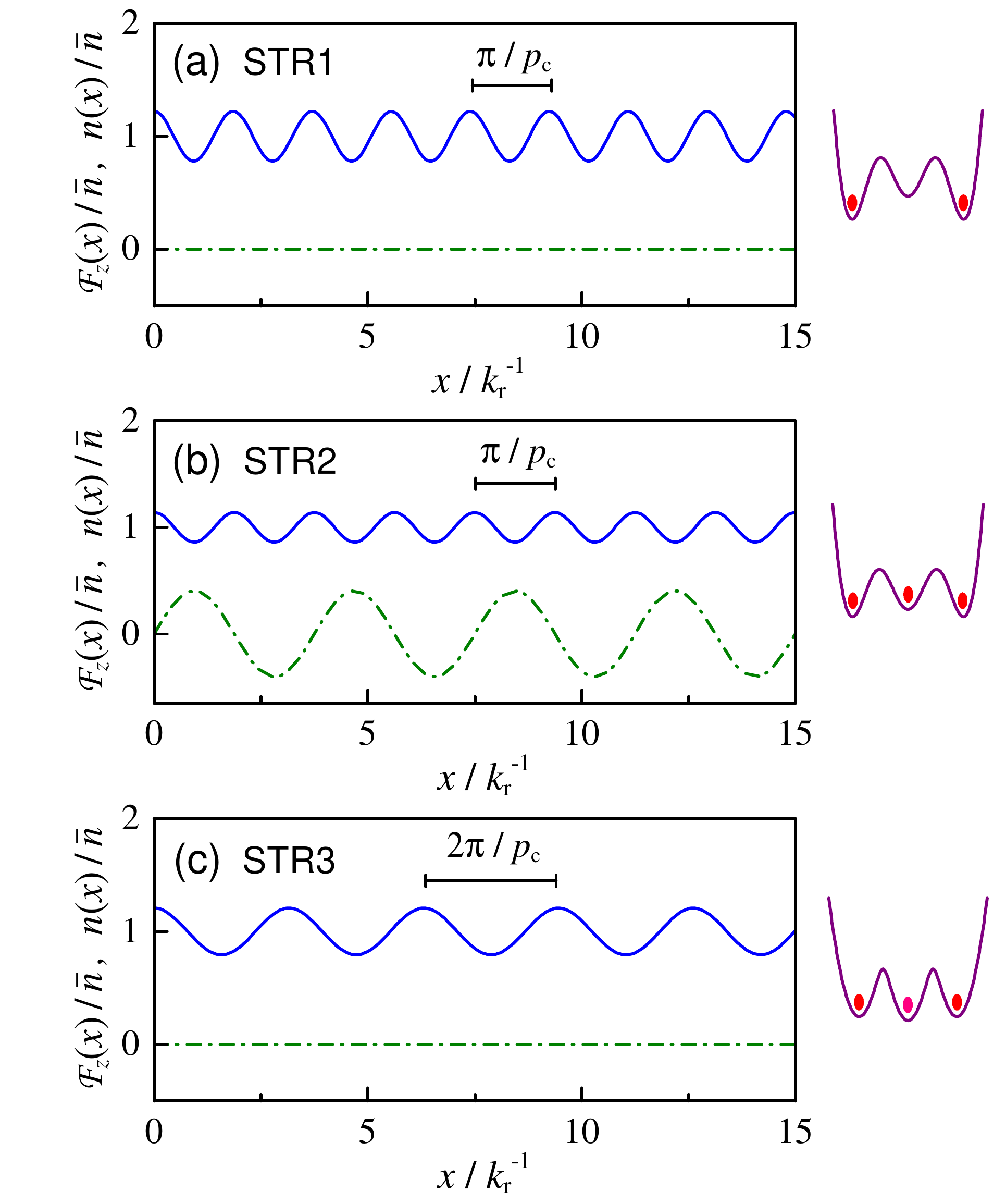}
\caption{Spatial distributions of density (solid line) and longitudinal magnetization (dash-dotted line) in (a) the STR1 phase, (b) the STR2 phase, and (c) the STR3 phase. Right column shows the schematic single-particle dispersion and the momentum occupation of the condensate in each case. Parameters: (a) $\Omega=3.6E_{\rm r}$, $\Lambda=-0.72E_{\rm r}$, $c_2\bar n=0.1E_{\rm r}$, (b) $\Omega=3.6E_{\rm r}$, $\Lambda=-0.68E_{\rm r}$, $c_2\bar n=0.1E_{\rm r}$, and (c) $\Omega=0.9E_{\rm r}$, $\Lambda=0$, $c_2\bar n=-0.1E_{\rm r}$; for all the plots, $c_0\bar n=1E_{\rm r}$.}
\end{figure}

In Fig.~6, we show examples of the density and magnetization distributions in the different stripe phases. It is worth noting that although the condensate wavefunction~(\ref{variation_wf}) is written in the rotating frame, our results for $n(\vec r)$ and $\mathcal{F}_z(\vec r)$ remain the same in the laboratory frame~\cite{note2}. These distribution functions can be experimentally measured through the {\it in situ} imaging.


For $c_2>0$, the STR1 phase extends a wide area in the phase diagram (see Fig.~1), and it is always the ground state when the quadratic Zeeman field $\Lambda$ is negative large enough. The STR2 phase appears in a narrow region, where the energies of the single-particle states at momenta $p_x=0$ and $\pm p_{\rm c}$ are close. The upper boundary and lower boundary of this region start at $(\Omega,\Lambda)=(0,0)$ and end at the tricritical point $(\Omega_\star,\Lambda_\star)$. For $c_2<0$, the STR3 phase exists when both $\Omega$ and $|\Lambda|$ are small, and it could extend to a larger $\Omega$ regime when the density-density interaction $c_0\bar n$ is suppressed. The area for the stripe phases in the phase diagram shrinks  as $|c_2|$ decreases, and it eventually vanishes when the interactions become spin-independent. In the limit that Raman coupling vanishes, our phase diagram recovers the well-known results of the spinor Bose gases~\cite{SpinorReview} (see Appendix~D for details).

We note that the phase region for the stripe condensates determined from the variational approach is qualitatively consistent with the instability analysis discussed in the previous section (see Fig.~1 and Fig.~5).

\subsection{Contrast of fringes and tricritical points}

To further characterize the stripe phases, we investigate the contrast of the density fringes, which is defined by the relative amplitude of the density modulation $(n^{\rm max}-n^{\rm min})/(2\bar n)$. The nonzero contrast in the stripe phases implies the spontaneous breaking of the translation symmetry and the emergence of the crystalline order. In the STR2 phase, besides the density modulation, the longitudinal magnetization also shows a spatial oscillation. As a result, the contrast of magnetization fringes $(\mathcal{F}_z^{\rm max}-\mathcal{F}_z^{\rm min})/(2\bar n)$ is nonzero. In the PW phase and ZM phase, both density and magnetization are uniform, and the crystalline order vanishes.

\begin{figure}[t!] 
\includegraphics[width=7.7cm]{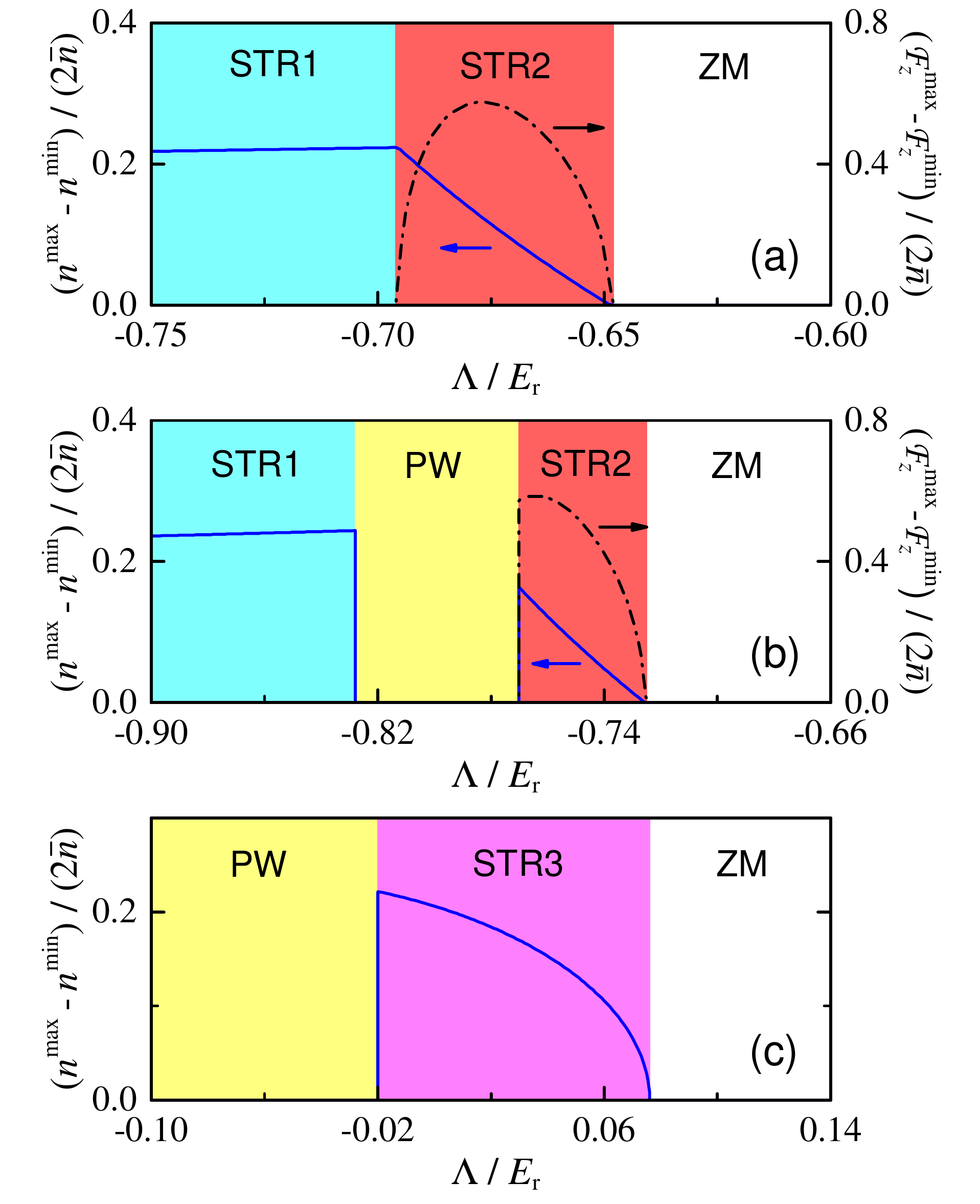}
\caption{Contrasts of density modulation (solid line) and magnetization modulation (dash-dotted line) as a function of $\Lambda$. Parameters: (a) $c_2\bar n=0.1E_{\rm r}$, $\Omega=3.6E_{\rm r}$, (b) $c_2\bar n=0.1E_{\rm r}$, $\Omega=3.85E_{\rm r}$, and (c) $c_2\bar n=-0.1E_{\rm r}$, $\Omega=0.9E_{\rm r}$. For all the plots, $c_0\bar n=1E_{\rm r}$. }
\end{figure}

The behaviors of contrasts across the different phase boundaries are shown in Fig.~7. The transitions between the PW phase and the stripe phases (STR1, STR2 and STR3) are first order. The contrast of density modulation exhibits a sudden jump at the phase boundary, and the magnetization per particle $M_z=\frac{1}{N} \int d\vec r \mathcal{F}_z$ also shows a discontinuity. The ZM-STR2 transition and the ZM-STR3 transition have the nature of second order. When either of these transitions takes place, the uniform condensate in the ZM phase evolves smoothly into a density modulated stripe with a finite contrast. For $c_2>0$, there is also an intriguing transition between two stripe phases (STR1 and STR2). This transition is second order as well. As the system enters the STR1 phase from the STR2 side, the coefficient $\xi_0$ in the wavefunction continuously decreases to zero, and the contrast of the magnetization modulation vanishes at phase boundary. It is worth noting that in the STR2 phase the contrast of magnetization fringes exhibits a nonmonotonic behavior as $\Lambda$ or $\Omega$ changes.

The first-order transitions and the second-order transitions meet at the tricritical points, as shown in the phase diagram of Fig.~1. For $c_2>0$, the tricritical point denoted by the symbol $\odot$ separates the STR1, STR2, and ZM phases, and the tricritical point denoted by the symbol $\bigstar$ separates the STR2, ZM, and PW phases. For $c_2<0$, the STR3, PW, and ZM phases are separated by the tricritical point denoted by the symbol $\bigtriangleup$, and the tricritical transition at $(\Omega_\star,\Lambda_\star)$ is essentially the same as the noninteracting case. As the spin-dependent interaction vanishes, the region for the stripe phases collapses, and the tricritical points $\odot$ and $\bigtriangleup$ disappear.

\section{Discussion and Conclusion}

Finally, we discuss the experimental relevance of our theory.

In the recent experiment of Ref.~\cite{NIST2015}, the synthetic SO coupling has been realized in the spin-1 Bose gas of $^{87}$Rb atoms. Due to the very weak ferromagnetic interaction ($c_2/c_0\simeq -0.005$), the stripe phase (STR3) in the $^{87}$Rb gas is expected to be found in a very limited region, and the contrast of the density modulation is too weak to make an evident detection. The interactions effects on the PW-ZM transition are also negligible; the experimental observation of the magnetic phase transition~\cite{NIST2015} is indeed in a good agreement with the noninteracting prediction.

Atomic gas of $^{23}$Na could be a candidate to observe the stripe phases with antiferromagnetic interaction. Theoretically, a STR2 condensate with high contrast of density and magnetization modulations can be achieved at suitable values of $\Omega$ and $\Lambda$. However, the narrow phase region requires the fine-tuned parameters in experiment. The STR1 phase, which extends a large area in the phase diagram, is more promising for an accessible experimental detection. To estimate the visibility of the density fringes, it is helpful to derive some analytic results when the quadratic Zeeman field is negative large. In the limit of $-\Lambda\gg\Omega,E_{\rm r}$, the hyperfine level $m_{\rm F}=0$ can be adiabatically eliminated, and the system can be mapped to a spin-half model with the effective Hamiltonian given by
\begin{align}
  H' = & \int d\vec r\,  \bigg[ \psi'^\dag \frac{\vec p^2}{2m} \psi' - \frac{2k_{\rm r}}{m} \psi'^\dag p_x\sigma_z \psi' +\frac{\Omega'}{2} \psi'^\dag \sigma_x \psi'
  \nonumber \\
  & \; + \frac{1}{2} \sum_{\sigma=\uparrow,\downarrow} g_{\sigma\sigma} \psi_{\sigma}'^\dag \psi_{\sigma}'^\dag \psi_{\sigma}' \psi_{\sigma}' + g_{\uparrow\downarrow} \psi_{\uparrow}'^\dag \psi_{\downarrow}'^\dag \psi_{\downarrow}' \psi_{\uparrow}' \bigg],
  \label{spin-half_model}
\end{align}
where $\psi'^\dag = (\psi_\uparrow'^\dag, \psi_\downarrow'^\dag)$ is the field operator for the spin-half atoms, $\uparrow$ and $\downarrow$ represent the hyperfine states $m_{\rm F}=1$ and $-1$, respectively, $g_{\uparrow\uparrow}=g_{\downarrow\downarrow}=c_0+c_2$ and $g_{\uparrow\downarrow}=c_0-c_2$ are interactions parameters, and $\Omega'=\Omega^2/(2\Lambda')$ is the effective coupling strength. From the knowledge of the spin-half model~\cite{Trento1}, it is readily to show the contrast of the stripe can reach a maximum value $\sqrt{2c_2/(c_0+2c_2)}/[1+c_0\bar n/(4E_r)]$. Thus, for sodium $(c_2/c_0\simeq0.03)$, the amplitude of density modulation can easily exceed 20\%. We have numerically checked that when the system is away from the negative large $\Lambda$ limit, a comparable contrast can also be achieved.

It should be noted that the configuration of hyperfine states in our spin-half model is different from the well studied two-component system of $^{87}$Rb atoms~\cite{NIST2011,USTC2012,NIST2013,USTC2014,WSU2014-1,WSU2014-2, USTC2015}. In that setup, two adjacent hyperfine levels, either $m_{\rm F}=0,+1$ or $m_{\rm F}=0,-1$,  are labeled as the pseudospin states, and the stripe phase exists only when $c_2<0$~\cite{note4}. For $^{87}$Rb, the maximum amplitude of the density modulation is about $4\%$. Such a weak contrast makes it very difficult to directly observe the density fringes.

The major challenge for the experiment with $^{23}$Na atoms may be the heating problem. Due to the small fine-structure splitting, the heating by the Raman lasers could be much more serious than in rubidium gases. For the spin-half system, it has been shown that the phase region for the stripe state is almost unchanged when temperature is below 0.5 $T_{\rm c}$~\cite{USTC2014,Yu2014} ($T_{\rm c}$ is the condensation temperature). For the case of spin-1, a complete phase diagram at finite temperature is still unavailable, and we leave this issue to future study.

In conclusion, the interplay between the SO coupling and spin-dependent interaction could give rise to a rich ground-state phase diagram in spin-1 Bose gases. Three types of stripe condensation states with different crystalline orders are identified in the presence of either antiferromagnetic or ferromagnetic interaction, and the occurrences of various tricritical points are predicted. Our results could be useful to the future exploration of the stripe condensates in experiment. \\

{\it Note added $-$} During the preparation of the paper we became aware of two recent works~\cite{ZhangCW2015,Giovanni2015}, in which a similar problem is also studied.

\begin{acknowledgments}
  The author would like to acknowledge helpful discussions with Hui Zhai, Chuanwei Zhang and Giovanni Martone.
\end{acknowledgments}

\appendix

\section{Series expansion of single-particle dispersion}

The lowest band of single-particle energy spectrum can be expanded as Eq.~(\ref{Ep_expansion}) with the coefficients given by
\begin{align*}
  \lambda_2 &= \frac{1}{2m} \left[ 1 + 16 E_{\rm r} \left( \frac{1}{D} + \frac{\Lambda'-D}{\Omega^2}\right) \right], \\
  \lambda_4 &= \frac{64E_{\rm r}^3}{m^2} \left[ \frac{1}{D^3} + \frac{4}{\Omega^2D} - \frac{2(\Lambda'+D)}{\Omega^4} - \frac{4\Lambda'^2(\Lambda'-D)}{\Omega^6} \right], \\
  \lambda_6 &= \frac{1024E_{\rm r}^3}{m^3} \left[ \frac{1}{D^5} +  \frac{3}{\Omega^2D^3} + \frac{18}{\Omega^4D} + \frac{2(\Lambda'-5D)}{\Omega^6} \right.
  \\
  & \qquad \qquad \quad \left. + \frac{16\Lambda'^3}{\Omega^8}
  + \frac{16\Lambda'^4(\Lambda'-D)}{\Omega^{10}}\right],
\end{align*}
where $D=\sqrt{\Lambda'^2+2\Omega^2}$ is the energy gap between the lowest band and highest band at $p=0$. When $\lambda_6>0$, the power expansion form of the dispersion relation~(\ref{Ep_expansion}) can be used to determine the PW-ZM transition.

\section{Self-Energy in Bogoliubov Approximation}

In Bogoliubov approximation, the self-energies $\Sigma_{\rm N}$ and $\Sigma_{\rm A}$ are independent of momentum and frequency. The matrix elements of the normal self-energy are given by
\begin{align}
  \Sigma_{\rm N}^{11} &= c_0 \bar n(1+\alpha_+^2) + c_2 \bar n (1+M_z- \alpha_-^2),
  \\
  \Sigma_{\rm N}^{22} &= c_0\bar n(1+\alpha_0^2) + c_2 \bar n(1-\alpha_0^2),
  \\
  \Sigma_{\rm N}^{33} &= c_0\bar n(1+\alpha_-^2) + c_2 \bar n(1-M_z-\alpha_+^2),
  \\
  \Sigma_{\rm N}^{12} &= \Sigma_{\rm N}^{21} = c_0\bar n\alpha_0\alpha_+ + c_2\bar n\alpha_0(\alpha_++2\alpha_-),
  \\
  \Sigma_{\rm N}^{23} &= \Sigma_{\rm N}^{32} = c_0\bar n \alpha_0\alpha_- + c_2\bar n\alpha_0(\alpha_-+2\alpha_+),
  \\
  \Sigma_{\rm N}^{13} &= \Sigma_{\rm N}^{31} = (c_0-c_2)\bar n\alpha_+\alpha_-,
\end{align}
where the spinor parameters $\alpha_\pm$ and $\alpha_0$ are obtained from the GP equation (\ref{GP_equation}). The matrix elements of the anomalous self-energy are given by
\begin{align}
  \Sigma_{\rm A}^{11} &= (c_0+c_2)\bar n\alpha_+^2,
  \\
  \Sigma_{\rm A}^{22} &= c_0 \bar n\alpha_0^2 + 2c_2 \bar n\alpha_+\alpha_-,
  \\
  \Sigma_{\rm A}^{33} &= (c_0+c_2)\bar n\alpha_-^2,
  \\
  \Sigma_{\rm A}^{12} &= \Sigma_{\rm A}^{21} = (c_0+c_2)\bar n\alpha_0\alpha_+,
  \\
  \Sigma_{\rm A}^{23} &= \Sigma_{\rm A}^{32} = (c_0+c_2)\bar n\alpha_0\alpha_-,
  \\
  \Sigma_{\rm A}^{13} &= \Sigma_{\rm A}^{31} = c_0\bar n\alpha_+\alpha_- + c_2\bar n(\alpha_0^2-\alpha_+\alpha_-).
\end{align}

Using the self-energies $\Sigma_{\rm N}$ and $\Sigma_{\rm A}$, the GP equation (\ref{GP_equation}) can be rewritten as
\begin{align}
  \big[ H_{\rm R}(p_{\rm c}) -\mu + \Sigma_{\rm N} - \Sigma_{\rm A} \big]
  \begin{pmatrix}
    \alpha_+ \\ \alpha_0 \\ \alpha_-
  \end{pmatrix} = 0.
\end{align}
Thus the Hugenholtz-Pines relation (\ref{HP_relation}) is verified at Bogoliubov level.

\section{Bragg spectroscopy in PW phase and ZM Phase}

\begin{figure} 
\includegraphics[width=7.4cm]{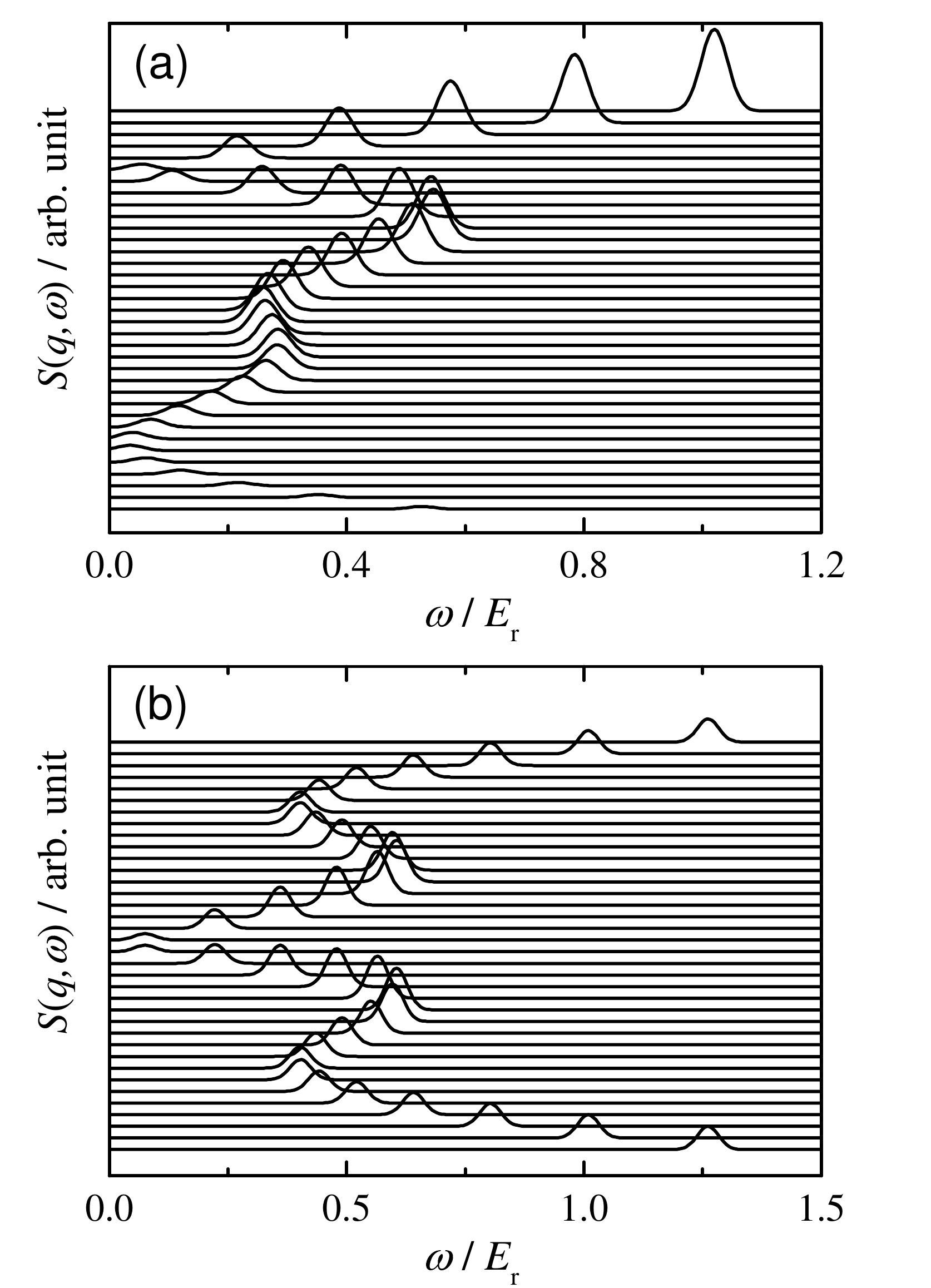}
\caption{Dynamic structure factor in (a) the PW phase and (b) the ZM phase. Each line is for a given wave vector $\vec q = q_x\hat x$. From bottom to top, the values of $q_x$ ranges from $-4.3k_{\rm r}$ to $1.8k_{\rm r}$ with an even spacing $0.15k_{\rm r}$ in (a) and  ranges from $-2.8k_{\rm r}$ to $2.8k_{\rm r}$ with an even spacing  $0.16k_{\rm r}$ in (b). Other parameters: (a) $\Lambda=-0.5E_{\rm r}$; (b) $\Lambda=-0.2E_{\rm r}$; for both plots, $c_0\bar n=1E_{\rm r}$, $c_2=0$, and $\Omega=3E_{\rm r}$.  }
\end{figure}

In the zero temperature limit, the Bragg spectrum measured in experiment is proportional to the dynamic structure factor~\cite{StringariBook},
\begin{align}
  S(\vec q,\omega)= \sum_{\ell} \left| \langle\Phi_\ell| \rho_{\vec q}^\dag | \Phi_0 \rangle \right|^2 \delta(\omega-\omega_{\ell0}),
  \label{structure_factor}
\end{align}
where $\rho_{\vec q}=\int d\vec r\, \psi^\dag \psi e^{-i\vec q\cdot \vec r}$ is the density fluctuation operator, $\Phi_0$ is the many-body ground state with the energy $E_0$, $\Phi_\ell$ is the excited state with the energy $E_\ell$, and $\omega_{\ell0}=E_{\ell}-E_0$.
In Bogoliubov approximation, $\rho_{\vec q}\simeq \int d\vec r\, (\varphi^\dag \psi + \psi^\dag \varphi)e^{-i\vec q\cdot \vec r}$, and $S(q,\omega)$ can be readily obtained by solving the quadratic Hamiltonian (\ref{H_Bogoliubov}).

In Fig.~8, we plot the dynamic structure factor for different wave vector $\vec q$ lying on the $x$-axis. To take account of the energy resolution in experiment, $\delta$-function in (\ref{structure_factor}) has been replaced by $\frac{1}{\sqrt{\pi}\epsilon}e^{-(\omega-\omega_{\ell0})^2/\epsilon^2}$ in our numeric calculation (set $\epsilon=0.03E_{\rm r}$). One can see the asymmetric (symmetric) double-roton structure in the PW (ZM) phase.

\section{Spinor wavefunction in vanishing Raman Coupling Limit}

As the Raman coupling is gradually switched off, our phase diagram recovers the well-known result in spinor Bose gases~\cite{SpinorReview}. To show this, it is more convenient to use the laboratory frame, in which the condensate wavefunction is given by $\tilde \varphi=e^{-2ik_{\rm r}xF_z}\varphi$, with $\varphi$ being the wavefunction in the rotating frame.

For $c_2>0$, the antiferromagnetic interaction favors the vanishing magnitude of $\mathcal{\vec F}$. In the limit that the Raman coupling vanishes, the ground state is the ZM phase when $\Lambda>0$ and is the STR1 phase when $\Lambda<0$. The ZM phase recovers the longitudinal polar state,
\begin{align}
  \tilde \varphi_{\rm ZM} & \xrightarrow{\Omega\rightarrow 0} \sqrt{\bar n} \begin{pmatrix}
    0 \\ e^{i\chi} \\ 0  \end{pmatrix},
\end{align}
and the STR1 phase recovers the transverse polar state,
\begin{align}
  \tilde \varphi_{\rm STR1} & \xrightarrow{\Omega\rightarrow 0} \sqrt{\frac{\bar n}{2}} \begin{pmatrix}
    1 \\ 0 \\ e^{i\chi} \end{pmatrix},
\end{align}
where $\chi$ is an arbitrary real number. A complicated situation occurs at $\Lambda=0$, where the STR1 phase, the STR2 phase and the ZM phase are degenerate. The STR2 phase also approaches to a specific type of polar state in the limit that both $\Omega$ and $\Lambda$ vanish,
\begin{align}
  \tilde \varphi_{\rm STR2} \xrightarrow{(\Omega,\Lambda)\rightarrow (0,0)} \sqrt{\frac{\bar n}{2}} \begin{pmatrix}
    \sqrt{1-|\xi_0|^2}e^{i\chi} \\ i\sqrt{2}|\xi_0| \\ \sqrt{1-|\xi_0|^2} e^{-i\chi}
  \end{pmatrix},
\end{align}
with $0<|\xi_0|<1$.

For $c_2<0$, the spin-dependent interaction prefers to generate a ferromagnetic order. In the $\Omega\rightarrow 0$ limit, the ground state is the PW phase when $\Lambda<0$ and is the STR3 phase when $0<\Lambda<2|c_2|\bar n$. The PW phase recovers to the longitudinal ferromagnetic state,
\begin{align}
   \tilde \varphi_{\rm PW} & \xrightarrow{\Omega\rightarrow 0} \sqrt{\bar n} \begin{pmatrix}
    e^{i\chi} \\ 0 \\ 0  \end{pmatrix},
\end{align}
and the STR3 phase recovers the partially magnetic state,
\begin{align}
  \tilde \varphi_{\rm STR3} & \xrightarrow{\Omega\rightarrow 0} \frac{\sqrt{\bar n}}{2} \begin{pmatrix}
   \sqrt{1+\frac{\Lambda}{2c_2\bar n}} e^{i\chi} \\ \sqrt{2-\frac{\Lambda}{c_2\bar n} } \\ \sqrt{1+\frac{\Lambda}{2c_2\bar n}} e^{-i\chi} \end{pmatrix}.
\end{align}
When $\Lambda>2|c_2|\bar n$, the ground state is the ZM phase. It recovers the longitudinal polar state in the same way as the case of $c_2>0$.


\begin{thebibliography}{99}
\bibitem{note1} We note that our notation for the Raman coupling strength is the same as the previous theoretical work in~\cite{Lan2014}. It is different from the experimental paper of Ref.~\cite{NIST2015} by a factor of $\sqrt{2}$.
\bibitem{note_singleparticle} While the single-particle ground state could be a superposition of the degenerate mometum states with an arbitrary weight, our results still make sense by assuming a vanishing small linear Zeeman field.
\bibitem{note_transition} In this calculation, the stripe phases are not considered. For the parameters used in Fig.~3, the STR2 phase appears in a very narrow region which can be ignored.
\bibitem{note2} In contrast, the distribution functions $\mathcal{F}_x$ and $\mathcal{F}_y$ show different spatial modulations in the labratory frame and the rotating frame. We do not discuss them in this work.
\bibitem{note4} In the spin-half model, the stripe phase only exists for $g_{\uparrow\uparrow}+g_{\downarrow\downarrow}>2g_{\uparrow\downarrow}$ \cite{Trento1}. When the pseudospin states are constructed by the two adjacent hyperfine levels of spin-1 bosons, the effective interaction parameters are given by $g_{\uparrow\uparrow}+g_{\downarrow\downarrow}=2c_0+c_2$ and $g_{\uparrow\downarrow}=c_0+c_2$.
\end{thebibliography}
\end{document}